\documentclass[12pt]{article}
\usepackage{amsmath}
\usepackage{amssymb}
\usepackage{amsfonts}
\usepackage[english]{babel}
\usepackage[centerlast,small]{caption2}
\usepackage{graphicx}
\usepackage{array}

\newcommand{\htemp}{\frac{\alpha_s}{4\pi}\left(C_F-\frac{C_A}{2}\right)}
\newcommand{\absk}{\left\vert k^{2}\right\vert}
\newcommand{\ld}{\ln^2\frac{2E\omega}{\mu^2}}
\newcommand{\ls}{\ln\frac{2E\omega}{\mu^2}}
\newcommand{\bz}{\bar{z}}
\newcommand{\lqcd}{\Lambda_{\mathrm{QCD}}}

\begin{document}

\begin{titlepage}

\begin{center}

{\bf \large Loop corrections to the form factors in $ B \to \pi l \nu$ decay}

\vspace{1cm}

G.G. Kirilin\footnote{G.G.Kirilin@inp.nsk.su}

\vspace{1cm}

Budker Institute of Nuclear Physics SB RAS\\
630090 Novosibirsk, Russia\\

\end{center}

\bigskip
\begin{abstract}

In this paper we study the semileptonic decay $B\to\pi l \nu$ and in particular
the factorizable contribution to symmetry breaking corrections to the form
factors at large recoil. This contribution is a convolution of the coefficient
function, which can be calculated in perturbation theory, and of the
nonperturbative light-cone distribution amplitudes of the mesons. The
coefficient function, in turn, can also be represented as convolution of the
hard Wilson coefficient and of the jet function. Loop corrections to the hard
Wilson coefficient and jet function are calculated. We use the method of
expanding by regions to calculate these corrections. The results obtained
coincide with the ones calculated in the framework of the soft-collinear
effective theory (SCET). Factorization of soft and collinear singularities into
the light-cone distribution amplitudes is demonstrated at one-loop level
explicitly. It is also demonstrated that the contribution of the so-called
soft-messenger modes vanishes; this fact is of critical importance to the
factorization approach to this decay.

\bigskip


\end{abstract}

\vspace{8cm}

\end{titlepage}

\section{Introduction}

Recent advances in experimental investigation of heavy  meson decays demand
detailed theoretical analysis of these decays. Of prime importance are
$B$-meson decays into light particles: pseudoscalar or vector mesons, photons
and light lepton pairs, because the amplitudes of these decays are proportional
to the non-diagonal elements of the CKM matrix.

Recently, consistent factorization approach has been developed for
heavy-to-light transitions; it includes so-called soft-collinear effective
theory (SCET) \cite{Bauer}, \cite{Beneke}. This theory describes form factors
of heavy-to-light transitions in the kinematical region with the energy of one
or several light final state particles comparable with the $B$-meson mass. A
typical example is the semileptonic decay $B\to \pi l \nu$. In first order in
$G_F$ the amplitude of the decay is proportional to the matrix element of the
vector current between $B$ and $\pi$ mesons. The matrix element can be written
in terms of two form factors:
\begin{multline}
A^{\mu}=\left\langle \pi(q)\right\vert \bar{q}\gamma^{\mu}Q\left\vert
B(p)\right\rangle \\=f_{+}(t)\,\left(  p^{\mu}+q^{\mu}-\frac{M_{B}^{2}-m_{\pi
}^{2}}{t}\,(p-q)^{\mu}\right)  +f_{0}(t)\,\frac{M_{B}^{2}-m_{\pi}^{2}}%
{t}(p-q)^{\mu}\,,\label{amplitude}%
\end{multline}
where $t=(p-q)^2$. Since $m_\pi\ll M_b$, we neglect the pion mass below.

\begin{figure}[tbp]
\begin{center}
\parbox[t]{4cm}{
\begin{center}
\includegraphics[width=4cm]{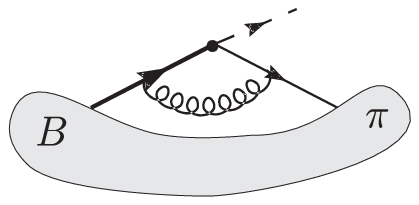}\\
\vspace{.3cm} (a) \end{center}} \hspace{2cm}
\parbox[t]{4cm}{
\begin{center}
\includegraphics[width=4cm]{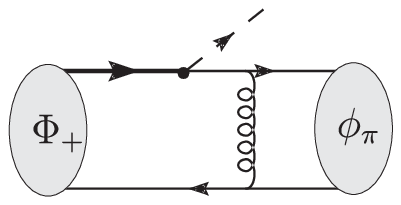}\\
\vspace{.3cm} (b) \end{center}} \caption{Symmetry breaking radiative
corrections}\label{ftree}
\end{center}
\end{figure}

Let us consider a frame where the momenta of the initial and final mesons take
the form:
\begin{align}
p=M_{B}~v=M_{B}~(1,0,0,0),\quad q=E~n=E(1,0,0,-1), \label{frame1}
\end{align}
where $E$ is the energy of the pion. It is convenient to introduce one more
light-cone vector:
\begin{align}
n_{+}=(1,0,0,1),\quad v=\frac{n}{2}+\frac{n_{+}}{2}.\label{frame2}%
\end{align}
In this frame the amplitude (\ref{amplitude}) can be represented as follows:
\begin{align}
A^{\mu}=\frac{M_{B}}{2}\left(  f_{0}+f_{+}~z\right)  ~n^{\mu}+\frac{M_{B}%
}{2~\bar{z}}\left(  f_{0}-f_{+}~z\right)  \,n_{+}^{\mu}\,,\label{amplitude2}%
\end{align}
where $z=2E/M_B, \bar{z}=1-z$. On the other hand, the matrix element can be
written as
\begin{align}
\left\langle \pi(q)\right\vert \bar{q}\gamma^{\mu}Q\left\vert B(p)\right\rangle
=\frac{n^{\mu}}{2}~\left\langle \pi(q)\right\vert
\bar{q}\,\not\!n_{+}\,Q\left\vert B(p)\right\rangle
+\frac{n_{+}^{\mu}}{2}\,\left\langle \pi(q)\right\vert
\bar{q}\,\not\!n\,Q\left\vert B(p)\right\rangle\,.\label{matrix}
\end{align}
The kinematical region we are interested in is $z\sim 1$, i.e., the energy of
the pion is of the order of the mass of the heavy meson $M_B \gg\lqcd$. If
hadronization occurs in such a way that the light quark produced in the decay
of the heavy quark inside the $B$-meson has a virtuality of order $\lqcd^2$ but
this quark carries sufficiently large part of the pion energy, then this quark
may be thought to be asymptotically free so that $\bar{q}\!\not\!\!n\approx 0$.
In this case the last term in (\ref{matrix}) vanishes. Comparing with
(\ref{amplitude2}) yields
\begin{align}
z f_+(t)=f_0(t). \label{symm}
\end{align}
Expression (\ref{symm}) is known as \textit{the symmetry of the form factors at
large recoil} \cite{Charles}. Apparently, a hard radiative correction, that
affects the light quark, breaks this symmetry. In \cite{Beneke0} possible ways
of violating the symmetry were considered, namely, the radiative corrections to
the "non-factorizable"\ Feynman mechanism (Fig.~\ref{ftree}a) and the hard
spectator-scattering mechanism (Fig.~\ref{ftree}b). The contribution of the
latter factorizes into the product of the light-cone distribution amplitudes
(LCDAs) of the mesons and the coefficient function. The difference of the form
factors can be written \cite{Stewart}, \cite{Beneke2}
\begin{multline} f_{0}-z f_{+}\,=\frac{\bar{z}}{M_{B}}\,\left\langle \pi\right\vert
\bar{q}\not\!\!n\,Q\left\vert
B\right\rangle=\mathcal{C}(E)\,\xi_{\pi}(E)+\int_{0}^{\infty
}d\omega\,\Phi_{+}(\omega)\int_{0}^{1}du\,\phi_{\pi}(u)\,T(E,\omega,u),
\label{factor}
\end{multline}
where $\xi_{\pi}(E)$ is the universal nonperturbative form factor that
corresponds to the Feynman mechanism, $\Phi^+(\omega),\,\phi_{\pi}(u)$ are the
light-cone distribution amplitudes of the initial and final mesons. The form
factors can be also represented in the form of Eq.\ref{factor} \cite{Neubert1},
\cite{Beneke2} but in this case the Wilson coefficient $\mathcal{C}(E)$ begins
with $1$ instead of $\alpha_s$ as it does with the difference of the form
factors~(Fig.~\ref{ftree}a).

The coefficient function $T(E,\omega,u)$ is universal in the sense that it
depends on initial and final states through the quantum numbers only.
Consequently, this function can be calculated by considering the scattering
amplitude of the partons that constitute the final and initial mesons:
\begin{align}
A(E,\omega,u)=\int d^{4}xe^{i(p-q)x}\left\langle \bar{q}(-q_{2}),q(q_{1}%
)\left\vert ~\bar{q}(x)\not\!n\,b(x)~\right\vert
\bar{q}(-k),b(p-k)\right\rangle\,,\label{defamp}
\end{align}
so that the fractions of the pion's longitudinal momentum carried by the
final-state light quark and antiquark are taken to be $q_1\approx \bar{u} q$
and $q_2\approx u q$, respectively (see Fig.~\ref{kinematics}), and $k\cdot n
=\omega$ (the choice of the external kinematics will be considered in detail in
Section 2). The coefficient function $T(E,\omega,u)$ is the amplitude
(\ref{defamp}) averaged over the spin and color states of the initial and final
mesons
\begin{align}
T(E,\omega,u)=K\,\bar{A}(E,\omega,u)\,,\label{cfunction}
\end{align}
where $K$ is a normalization factor, which depends on normalizations of the
light-cone distribution amplitudes. For averaging over the spin and color
states we use the following projection operators:
\begin{align}
q_{\beta}^{j}(-q_{2})\,\bar{q}_{\alpha}^{i}(q_{1})  & \rightarrow K_{\pi}%
\,\delta^{ij}\not\!n_{\alpha\beta}\,,\label{p1}\\
Q_{\beta}^{j}(p-k)\,\bar{q}_{\alpha}^{i}(-k)  & \rightarrow K_{B}\,\delta
^{ij}\left[  \frac{(1+\not\!v)}{2}\frac{\not\!n \not\!n_{+}}{4}\right]
_{\alpha\beta}\,,\label{p2}
\end{align}
where Greek letters denote Lorentz indices and Latin letters denote color ones.
$K_\pi$ and $K_B$ are products of all normalization factors for the LCDAs.

Taking into account all mentioned above, one can calculate the contribution of
the diagram~\ref{ftree}b
\begin{align}
A_{1b}=-\frac{g^{2}d_{\mu\nu}}{(q_{1}-k)^{2}}\,\left[  \bar{q}(q_{1}%
)\,\gamma^{\mu}\,t^{a}\frac{(\not\!q-\not\!k)}{(q-k)^2}\not\!n\,Q(p-k)\right]
\left[ \bar {q}(-k)\,\gamma^{\nu}t^{a}q(-q_{2})\right].\label{a1b}
\end{align}
The substitution of the projectors (\ref{p1}) and (\ref{p2}) into (\ref{a1b})
yields
\begin{align}
A_{1b}  & \approx K_{\pi}\,K_{B}\,C_{F}\,N_{c}\,\frac{g^{2}d_{\mu\nu}}%
{2E^{2}u\omega}\,\mathrm{Sp}\left\{  \not\!n\,\gamma^{\mu}\,\frac{\not\!n_{+}\!\not\!n%
}{4}\frac{(1+\not\!v)}{2}\frac{\not\!n\not\!n_{+}}{4}~\gamma^{\nu}\right\} \label{sp1}\\
& =-K_{\pi}\,K_{B}\,C_{F}\,N_{c}\,(D-2)\,\frac{g^{2}}{E^{2}\,\omega\,
u}\,,\label{sp2}%
\end{align}
where $D$ is the dimensionality of space-time. It is convenient to define the
normalization factor $K$ as
\begin{align}
K^{-1}=- 4\pi\,K_{\pi}\,K_{B}\,C_{F}\,N_{c}\,(D-2)\,.
\end{align}
Therefore, the tree-level  coefficient function is
\begin{align}
T(E,\omega,u)=\frac{\alpha_s}{E^2\,\omega\,u}\,.
\end{align}
Note that even before the calculation of the trace (\ref{sp1}) one can see that
the gluon can only have the polarization orthogonal to the $(n, n_+)$ plane:
\begin{align}
d^{\mu\nu}\to d_\perp^{\mu\nu}=\delta^{\mu\nu}-\frac{n^\mu n_+^\nu+n^\nu
n_+^\mu}{2}\,. \label{polarization}
\end{align}
Putting it another way, there is an exchange of a transverse gluon in the
diagram \ref{ftree}b, and the factor $(D-2)$ in the expression (\ref{sp2}) is
the number of possible polarization states of the gluon. It is consistent with
the fact that the factorizable contribution can be generated by the operators
of $\mathrm{SCET_I}$ with $D_\perp$ only  \cite{Stewart,Beneke2,Neubert1}.

Calculation of loop corrections to the tree-level coefficient function is given
in the following Sections. In contrast to SCET, we calculate QCD diagrams
directly using \textit{the method of expanding by regions}. We outline this
method in the next Section. In Section 3 we present the general structure of
the coefficient function. Results and conclusions are given in Sections~4
and~5.


\section{The method of expanding by regions}

 In this Section we briefly review \textit{the method of expanding by
regions} (for further information the reader is referred to \cite{Smirnov1},
\cite{Beneke2}). Let us assume that there is a small parameter in a set of
one-loop diagrams that is a ratio of kinematical invariants composed by
external momenta $p_n$, such as $\displaystyle\left(p_n\cdot p_m\right)$. In
this case the method of expanding by regions allows one to calculate the result
for each diagram up to power suppressed terms with respect to the small
parameter. The idea is to single out the regions containing leading logarithms
from the entire integration domain in the loop momentum. That is a non-trivial
step, the point is to find field degrees of freedom that are most important for
the given process. Using factorization scales $\mu_i$ one has to divide the
loop integration domain into regions so that a typical loop momentum is
considered to be of the order of one of the scales specified by external
kinematics. After expansion of the integrands in every given region it is
necessary to regularize the integrands in covariant way so that the integral
should be convergent not only in the respective region but in the entire
integration domain. Usually dimensional regularization is
sufficient\footnote{See however \cite{Beneke2}.}. The last step is the
integration over the full loop momentum space. Thus the scales of the bounds of
the regions $\mu_i$ will occur in the form of $\displaystyle\ln \mu_i/\mu_j$
and/or $\displaystyle\ln \mu_i/\left(p_n\cdot p_m\right)$ (In addition, there
are poles in $\epsilon=(4-D)/2$ in the dimensional regularization). All the
intermediate scales $\mu_i$, by means of which separation into the regions has
been performed, cancel out in the sum of the contributions of all the regions.
Therefore, one can omit the contributions that contain only logarithms of the
ratio of the intermediate scales $\mu_i$ from the outset. It corresponds to
omitting scaleless integrals in the dimensional regularization.

\begin{figure}[tbp]
\begin{center}
\parbox[c]{6cm}{
\begin{center}
\includegraphics[width=6cm]{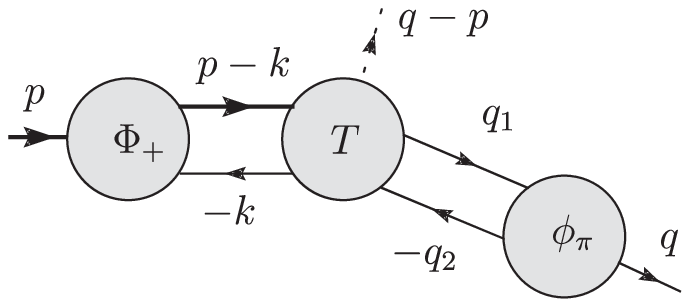}
\end{center}}
\hspace{2cm}
\parbox[c]{6cm}{
\begin{center}
\includegraphics[width=5cm]{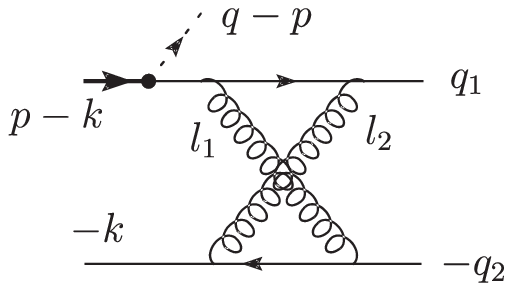}
\end{center}}\\
\parbox[c]{6cm}{
\begin{center}
\caption{Choice of kinematics}\label{kinematics}
\end{center}}
\hspace{2cm}
\parbox[c]{6cm}{
\begin{center}
\caption{One-loop example}\label{example}
\end{center}}
\end{center}
\end{figure}

To calculate loop corrections to the difference of the form factors, we choose
the external kinematic in such a way so as to regularize infrared singularities
by one or another external momentum squared (see Fig.~\ref{kinematics}).

For this purpose, we take the rest frame of the $B$-meson (\ref{frame1}),
(\ref{frame2}) so that $p^2=m_b^2\approx M_B^2$. From the physical point of
view it is clear that all components of the light spectator are of order
$\lqcd$ but, for the sake of convenience, we choose the momentum $k^{\mu}$ to
be in the $(n, n_+)$ plane:
\begin{align}
k^{\mu}  & =\omega~\frac{n_{+}^{\mu}}{2}-\frac{\mid k^{2}\mid}{\omega}%
~\frac{n^{\mu}}{2}.%
\end{align}
The momenta of the pion constituents (the quark and antiquark) are taken to be
\begin{align}
q_{1}  & =\bar{u}~E~n^{\mu}+q_{\perp}\label{qperp1}\,,\\
q_{2}  & =u~E~n^{\mu}-q_{\perp}\label{qperp2}\,,%
\end{align}
where $u$ and $\bar{u}$ are the fractions of the pion momentum such that $u,
\bar{u}>0$, $u+\bar{u}=1$, and the vector $q_\perp$ is orthogonal to the $(n,
n_+)$ plane. The small parameter needed for the expansion is
$\lambda^2=\lqcd/m_b\ll 1$. The hierarchy of the external kinematical
parameters is stated to be $E\sim m_b \gg \omega \sim \sqrt{- k^2} \sim
\sqrt{q^2_\perp} \sim\lambda^2 m_b$.

With this choice of kinematics logarithmic contributions are generated by five
regions. The corresponding kinematical parameters and momentum scaling are
presented in Table \ref{tab1}. All dimensional quantities are given in units of
$m_b$.

\begin{table}[h]
\begin{center}
\setlength{\extrarowheight}{1mm}
\begin{tabular}
[c]{|l|l|c|}\hline
 & Scale & $
l=\left(  l\cdot n,~l_{\perp},~l\cdot n_{+}\right)
$\\\hline\hline
hard & $m_{b}^{2}\sim1$ & \multicolumn{1}{|c|}{$(1,1,1)$}\\\hline
hard-collinear & $2 q\cdot k=2 E\omega\sim\lambda^{2}$ &
$(\lambda^{2},\lambda,1)$\\\hline
collinear & $\left\vert q_{i}^{2}\right\vert =\mathbf{q}_{\perp}^{2}%
\sim\lambda^{4}$ & \multicolumn{1}{|c|}{$(\lambda^{4},\lambda^{2},1)$}\\\hline
soft & $\left\vert k^{2}\right\vert \sim\lambda^{4}$ &
$(\lambda^{2},\lambda^{2},\lambda^{2})$\\\hline
soft-collinear & $\mathbf{q}_{\perp}^{2}~\absk/(2 E\omega)%
\sim\lambda^{6}$ & $(\lambda^{4},\lambda^{3},\lambda
^{2})$\\\hline
\end{tabular}
\caption{}\label{tab1}
\end{center}
\end{table}

In order to demonstrate the method of expansion by regions we consider the
diagram shown in Figure \ref{example}. All the results of loop corrections will
be normalized to the tree-level amplitude (\ref{sp2}). The integration measure
is defined by
\begin{align}
[dl]= \left(\frac{\mu^2 e^{\gamma_E}}{4\pi}\right)^\epsilon \frac{d^D
l}{(2\pi)^D}\,,\qquad D=4-2\epsilon\,.
\end{align}
The heavy quark does not participate in the scattering in this diagram,
therefore, there is no contribution of the hard region. The contribution of the
other regions are the following:\\

\noindent\textit{Hard collinear region}.
\begin{align} A^{hc}_{\ref{example}}=\htemp\left(
\frac{\mu^{2}}{2E\omega}\right)
^{\epsilon} \left[  \frac{1}{\epsilon^{2}}\frac{\Gamma^{2}(1-\epsilon)}%
{\Gamma(1-2\epsilon)}-\frac{1}{\epsilon}\left(  \frac{\ln u}{\bar{u}}%
+2\ln\bar{u}\right)  +\frac{\ln^{2}u}{2\,\bar{u}}+\ln^{2}\bar{u}\right].
\end{align}

\noindent\textit{The first collinear region}. In this region the gluon with
momentum $l_2$ (see. Fig.~\ref{example}) is collinear and the gluon with
momentum $l_1$ is hard-collinear. The contribution of this region is

\begin{align}A^{c}_{\ref{example}}=\htemp\left(  \frac{\mu^{2}}{\mathbf{q}_{\perp}^{2}%
}\right)  ^{\epsilon}\left[  -\frac{1}{\epsilon^{2}}\frac{\Gamma
^{2}(1-\epsilon)}{\Gamma(1-2\epsilon)}+\frac{\ln u}{\epsilon~\bar{u}}%
+\frac{1}{\bar{u}}~\left(
\mathrm{Li}_2\left(-\frac{\bar{u}}{u}\right)-\mathrm{Li}_2(\bar{u}) \right)
\right].
\end{align}

\noindent\textit{The second collinear region}. In this region the gluon with
momentum $l_1$ (see. Fig.~\ref{example}) is collinear and the gluon with
momentum $l_2$ is hard-collinear. The contribution of this region takes the
form
\begin{align}A^{c}_{\ref{example}}=\htemp\left(  \frac{\mu^{2}}{\mathbf{q}_{\perp}^{2}%
}\right)  ^{\epsilon}\left[
\frac{\ln\bar{u}}{\epsilon}-\mathrm{Li}_2(u)+\mathrm{Li}_2\left(
-\frac{u}{\bar{u}}\right)  \right].
\end{align}

\noindent\textit{Soft region}.
\begin{align}A^{s}_{\ref{example}}=\htemp\left( \frac{\mu^{2}}{\absk}\right)
^{\epsilon}\left[ -\frac{\Gamma^{2}(1-\epsilon)}{\epsilon^{2}~\Gamma
(1-2\epsilon)}\right].
\end{align}

\noindent\textit{Soft-collinear region}.
\begin{align}A^{sc}_{\ref{example}}=\htemp\left(  \frac{\mu^{2}2E\omega}{\absk%
\mathbf{q}_{\perp}^{2}}\right)  ^{\epsilon}\left[  \frac{\Gamma
(1-\epsilon)~\Gamma(1+\epsilon)}{\epsilon^{2}}~+~\frac{\ln\bar{u}}{\epsilon
}+\frac{\ln^{2}\bar{u}}{2}\right].\label{sc}
\end{align}

\noindent\textit{The sum}. Adding up all the contributions yields
\begin{multline}
 A_{\ref{example}}=\htemp\left[\ln\frac{2E\omega}{\absk}\ln\frac{2E\bar{u}~\omega
}{\mathbf{q}_{\perp}^{2}}+\left(  2\ln\bar{u}+\frac{\ln u}{\bar{u}}\right)
~\ln\frac{2E\bar{u}~\omega}{\mathbf{q}_{\perp}^{2}}\right.\\
\left.-2~\mathrm{Li}_2(u)-\frac{2}{\bar{u}}~\mathrm{Li}_2(\bar{u})-\ln^{2}\bar{u}
+\frac{\pi^{2}}{3}\right].
\end{multline}

\noindent Since the diagram is ultraviolet and infrared convergent, the
logarithms of the factorization scale $\mu$ as well as the singularities in
$\epsilon$ cancel in the sum of all the contributions.

As it will be demonstrated below at one-loop level, the regions with $l^2\sim
\lqcd^2$ factorize from the hard and hard-collinear regions into the light-cone
distribution amplitudes of the mesons (see also \cite{Beneke2}). Our analysis
is in many ways similar to the analysis performed in \cite{Sachrajda} for the
form factors that parametrize the amplitude of the radiative decay $B\to V
\gamma$.

As shown in \cite{Beneke2}, the contributions of the hard and hard-collinear
regions are also factorized. Corresponding logarithms $\ln m_b/\mu$ and $\ln
2E\omega/\mu^2$ can be summed up into a hard Wilson coefficient and a so-called
jet function. Some heuristic considerations of this factorization are presented
in the following Section.


\section{Structure of the coefficient function}

\begin{figure}[tbp]
\begin{center}
\parbox[c]{6cm}{
\begin{center}
\includegraphics[width=5cm]{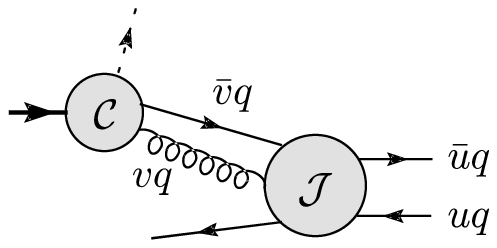}
\end{center}}
\hspace{2cm}
\parbox[c]{6cm}{
\begin{center}
\includegraphics[width=4.5cm]{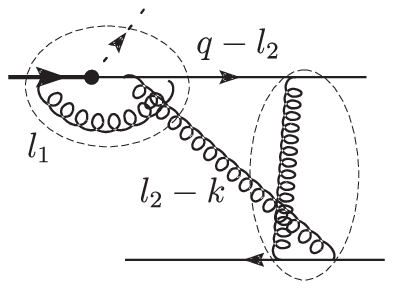}
\end{center}}\\
\parbox[c]{6cm}{
\begin{center}
\caption{Structure of the coefficient function}\label{sub}
\end{center}}
\hspace{2cm}
\parbox[c]{6cm}{
\begin{center}
\caption{Two loop example}\label{example2}
\end{center}}
\end{center}
\end{figure}

It is evident that the largest hard physical scale of the transition $B\to\pi$
is the heavy quark mass $m_b\sim M_B$. Therefore, the contribution of the hard
region is somehow or other associated with the heavy quark decay. However, the
hard-collinear scale $2E\omega$ implies participation of the light spectator
with the momentum  $k^\mu\sim \lqcd$ in the process. Consequently, there are
two subprocesses: the hard decay of the heavy quark into two partons that have
$v$ and $\bar{v}$ fractions of the collinear momentum $q$, and semihard
scattering of these partons by the soft spectator with the pion in final state
(see Figure \ref{sub}). To illustrate the general idea we consider an example
of a radiative correction to the diagram Fig.~\ref{example}. Namely, consider
the two-loop correction shown in Figure~\ref{example2} where the dash lines
contour the subprocesses involved. In the case when the momentum $l_1$ is hard
and $l_2$ is hard-collinear it is easy to calculate the contribution of this
diagram sequentially in three steps: the first step is integration over the
hard momentum $l_1$. In this step the virtualities of the quark and gluon
produced in the heavy quark decay can be neglected, i.e., the momenta of the
partons can be considered to be collinear to~$n$:

\begin{align}
\left(  q-l_{2}\right)  ^{\mu}\sim\left(  l_{2}-k\right)  ^{\mu}\sim
(\lambda^{2}, \lambda, 1) \Rightarrow\left\{
\begin{array}
[c]{c}%
\left(  l_{2}-k\right)^{\mu}\approx v~q^{\mu}\\
\left(  q-l_{2}\right)^{\mu}\approx\bar{v}~q^{\mu}%
\end{array}
\right.\,,
\end{align}
where $v=l_2\cdot n_+/(2E)$ is a dimensionless Sudakov parameter and
$\bar{v}=1-v$. This part contains logarithms of the hard scale $\ln m_b/\mu$,
we call it hard Wilson coefficient $C$ (hard coefficient function)
\begin{align}
C\left(v,\ln\frac{m_{b}}{\mu}%
,\frac{E}{m_{b}},\alpha_{s}\right)=\left\langle g (vq),q(\bar{v}q)\left\vert
~\bar{q}\not\!n\,b~\right\vert b(p-k)\right\rangle\,,
\end{align}
averaging over the spin and color states is performed in the same way as for
the amplitude (\ref{cfunction}). The second step is integration with respect to
the components of the momentum $l_2$ except $v$, i.e., over the components that
do not appear in the hard loop. This part, the so-called jet function, contains
the logarithms of the semihard scale $\ln 2E \omega/\mu^2$:
\begin{align}
J\left(  u,v,\ln\frac{2E\omega}{\mu^{2}%
},\alpha_{s}\right)=E^2\,\left\langle \bar{q}(-q_{2}),q(q_{1})\vert\,
g(vq),q(\bar{v}q),\bar{q}(-k)\right\rangle\,.
\end{align}
The last step is a convolution of the hard Wilson coefficient and the function
with respect to the Sudakov parameter $v$, this convolution is convergent,
i.e., it does not produce "big"\ logarithms:
\begin{align}
T(u,\omega,\mu,m_{b},E)=\frac{1}{E^2}\int_{0}^{1}dv~C\left(  v,\ln\frac{m_{b}}{\mu}%
,\frac{E}{m_{b}},\alpha_{s}\right)  ~J\left(  u,v,\ln\frac{2E\omega}{\mu^{2}%
},\alpha_{s}\right)\,.
\end{align}
In terms of effective theory the first step of our consideration corresponds to
the matching of the vector current in full QCD onto the set of three-body
operators of $\mathrm{SCET}_\mathrm{I}$ (integrating out of hard modes), the
second step corresponds to matching of the operators of the intermediate
effective theory $\mathrm{SCET}_\mathrm{I}$ onto four-quark operators of
$\mathrm{SCET}_\mathrm{II}$.

This simple partonic picture is only valid in the physical light-cone gauge
$A\cdot n_+=0$ because in this gauge only one transverse gluon connects $C$ to
$J$ but in the Feynman gauge it is accompanied by longitudinal collinear
gluons. To be more specific, in the light-cone gauge the hard-collinear gluon
produced in the heavy quark decay has a polarization that is approximately
transverse to the $(n, n_+)$ plane (see Eq.\ref{polarization})
\begin{align}
d_{\mu\nu}(l_2-k)=\delta^{\mu\nu}-\frac{(l_{2}-k)^{\mu}~n_{+}^{\nu}+(l_{2}-k)^{\nu}~n_{+}^{\mu}%
}{2~(l_{2}-k)\cdot n_{+}-i0}\approx d_{\perp}^{\mu\nu}\,.%
\end{align}
It is consistent with the fact that in this gauge the intermediate operator of
$\mathrm{SCET}_\mathrm{I}$ has the form (here we use notation of
\cite{Beneke2}):
\begin{align}
O_{1}^{1}(s_{1},s_{2})=\bar{\xi}_{C}(s_{1})
\not\!\!A_{\perp}(s_{2})\,h_{v}(0)\,.
\end{align}

As can be seen from the consideration of the diagram~\ref{ftree}b, the
tree-level hard Wilson coefficient and the jet-function are
\begin{align}
C\left(  v,\ln\frac{m_{b}}{\mu}%
,\frac{E}{m_{b}},\alpha_{s}\right)=1\,, \qquad J\left(  u,v,\ln\frac{2E\omega}{\mu^{2}%
},\alpha_{s}\right)=\frac{\alpha_s}{u\,\omega}\,\delta(u-v)\,.
\end{align}
The loop corrections to them are presented in the following Sections.


\section{Loop corrections}
Following are the contributions of the five regions. The contributions of each
diagram in all the regions except for the hard-collinear one are presented in
the Feynman gauge. The expressions for the hard-collinear region are given in
the light-cone gauge. As noted above, the loop corrections are normalized to
the tree-level amplitude (\ref{sp2}).

One further comment is in order. We use the "naive dimensional regularization"\
(NDR) scheme that corresponds to omitting $\gamma_5$ in the projection
operators (\ref{p1}) and (\ref{p2}). It can easily be shown that this scheme
(using the projection operators (\ref{p1}) and (\ref{p2}) and calculation of a
trace in dimensional regularization, i.e., without recourse to the Fierz
transformations) is identical to the scheme with the basis of the evanescent
operators chosen in \cite{Becher1}.

\subsection{Hard region}

\begin{figure}[tbp]
\begin{center}
\parbox[c]{3cm}{
\begin{center}
\includegraphics[width=3cm]{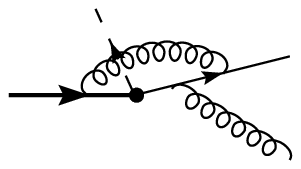}\\
(a) \end{center}} \hspace{2cm}
\parbox[c]{3cm}{
\begin{center}
\includegraphics[width=3cm]{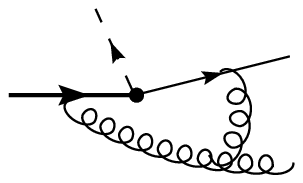}\\
(b) \end{center}}\hspace{2cm}
\parbox[c]{3cm}{
\begin{center}
\includegraphics[width=3cm]{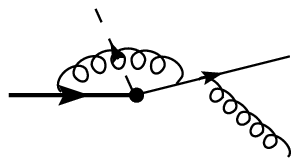}\\
(c) \end{center}}\newline
\vspace{.3cm}
\parbox[t]{3cm}{
\begin{center}
\includegraphics[width=3cm]{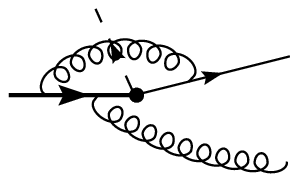}\\
(d) \end{center}} \hspace{2cm}
\parbox[t]{3cm}{
\begin{center}
\includegraphics[width=3cm]{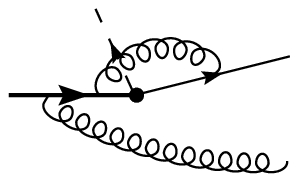}\\
(e) \end{center}} \hspace{2cm}
\parbox[t]{3cm}{
\begin{center}
\includegraphics[width=3cm]{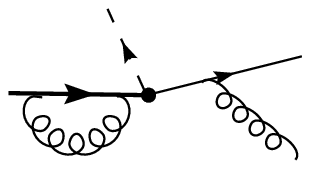}\\
(f) \end{center}}
\caption{Loop correction to the hard coefficent function}
\label{fHard}
\end{center}
\end{figure}

We start the discussion of loops with the hard region. As stated earlier, it is
necessary to consider the diagrams of the heavy quark decay into the collinear
quark and transverse gluon (see Fig.~\ref{fHard}). The contribution of each
diagram is the following:
\begin{equation}
C\left(u,z,\ln \frac{m_b}{\mu},\alpha_s\right)= 1 +
\frac{\alpha_s}{4\pi}\:\delta C\left(u,z,\ln \frac{m_b}{\mu}\right)\,,
\end{equation}
\begin{eqnarray}
\delta C_{a} & = &  \,\left( C_{F}-\frac{C_{A}}{2} \right) \,\left[ -\left(
2\ln ^{2}\frac{z\,m_b}{\mu }+\frac{\pi ^{2}}{12}+\frac{
2z}{\bz}\ln z+2\,\mathrm{Li}_{2}\left(\bz\right) \right) \right.  \nonumber\\
& & -\frac{2}{u}\,\left( 2\,\ln \bar{u}\,\left( \ln \,\frac{m_b}{\mu }+\ln z
-1\right) +\ln ^{2}\bar{u}+\mathrm{Li}_{2}\left( 1-\bar{u}\,z\right) -
\mathrm{Li}_{2}\left(\bz\right) \right)  \nonumber\\
& & \left. +\frac{1}{u}\,\left( \frac{\ln z}{\bz}-\frac{\ln\, \bar{u}z}{(1-
\bar{u}\,z)}+\ln \bar{u}\right) \right],\label{ehard1}\\
\delta C_{b} & = &  \, \left( -\frac{C_{A}}{2}\right) \, \left[ 2\ln
^{2}\frac{z m_b}{\mu }+\frac{\pi ^{2}}{12}+\frac{2\, z\, \ln z}{
\bz}+2\, \mathrm{Li}_{2}\left(\bz\right)\right.\nonumber\\
& &\left. -\frac{1}{\bar{u}}\, \left( \frac{
\ln z}{\bz}-\frac{\ln\, uz}{(1-u\, z)}-3\, \ln u\right) \right],\\
\delta C_{c} & = &  \, C_{F}\, \left[ -2\, \ln \frac{m_b}{
\mu }-\frac{1+\bz}{\bz}\, \ln z+1\right]\\
\delta C_{d} & = &  \, \left( C_{F}-\frac{C_{A}}{2}\right) \, \left[
-\frac{1}{\bar{u}}\, \left( \frac{\ln z}{\bz}-\frac{\ln\, u z}{
(1-u\, z)}+\ln u\right)\right.\nonumber\\
& &\left. -\frac{1}{u}\, \left( \frac{\ln z}{\bz}-\frac{
\ln \, \bar{u}z}{(1-\bar{u}\, z)}+\ln \bar{u}\right) \right],\\
\delta C_{e} & = &  \, C_{F}\, \left[ \frac{z}{\bz\, (1-u\,
z)}+\frac{1}{\bar{u}}\, \left( \frac{\ln z}{\bz^{2}}-\frac{\ln\, u z
}{(1-u z)^{2}}\right) +\frac{\ln u}{\bar{u}}\right],\\
\delta C_{f} & = &  \, C_{F}\, \left[ 3\, \ln \frac{m_b}{ \mu
}-2\right].\label{ehard2}
\end{eqnarray}
Note that the expressions (\ref{ehard1}--\ref{ehard2}) are finite in the limits
$u\rightarrow 0$ and $u\rightarrow 1$, it confirms the fact that  no logarithms
of $m_b/\mu$ or $2E\omega/\mu^2$ are generated by the convolution of the hard
Wilson coefficient and jet function. The final result of the hard Wilson
coefficient takes the form
\begin{eqnarray}
 \delta C & = & \,\left( C_{F}-\frac{C_{A}}{2}\right) \,
\left[ -\frac{2}{u}\,\left( 2\,\ln \bar{u}\,\left( \ln \,\frac{m_b}{\mu }+\ln z
-1\right) +\ln ^{2}\bar{u}+\mathrm{Li}_{2}\left( 1-\bar{u}\,z\right) -
\mathrm{Li}_{2}\left( \bz \right) \right) \right.  \nonumber\\
& &\left. -\frac{2}{\bar{u}}\,\left( \frac{\ln z}{\bz}-\frac{\ln u\,z}{
(1-u\,z)}-\ln u\right) \right]  \nonumber\\
& & +  \,C_{F}\,\left( -2\ln ^{2}\frac{z\,m_b}{\mu }-\frac{ \pi ^{2}}{12}+\ln
\frac{m_b}{\mu }+\frac{2+z}{\bz}\,\ln z-2\,\mathrm{Li}
_{2}\left(\bz\right) -1+\frac{z}{\bz\,(1-u\,z)}\right.  \nonumber\\
& & \left.+ \frac{1}{\bar{u}}\,\left( \frac{\ln z}{\bz^{2}}-\frac{\ln u\,z
}{(1-u\,z)^{2}}+\frac{\ln z}{\bz}-\frac{\ln u\,z}{(1-u\,z)}-2\,\ln u\right)
\right).\label{3BodCoef}
\end{eqnarray}
As already mentioned, this result is a special case of the Wilson coefficient
of the QCD vector current matched onto the corresponding three-body
$\mathrm{SCET}_{\mathrm{I}}$ operator. Our result coincides with the result
calculated in \cite{Neubert1} and the corresponding combination of matching
coefficients calculated in \cite{Beneke1} where an operator basis different
from \cite{Neubert1} is used.


\subsection{Hard-collinear region}

\begin{figure}
\begin{center}
\includegraphics[width=.3\textwidth]{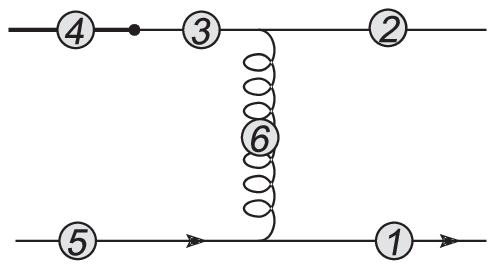}%
\caption{}\label{jnum}
\end{center}
\end{figure}
To calculate the contributions to the jet function, it is necessary to consider
all diagrams of the scattering of the hard-collinear quark and gluon by the
soft spectator.

In the diagram depicted in Fig. \ref{jnum} we label the external lines by (1),
(2), (4), (5) and the internal propagators by (3) and (6). The contributions to
the corrections to the jet function can be generated by diagrams in which a
gluon is attached to two labelled lines\footnote{There is quark loop
contribution to the diagram \{66\}}. We denote the contribution of each diagram
by the letter $J$ with $\{i j\}$ subscript where $i$ and $j$ are the numbers of
positions in Figure \ref{jnum}.

As emphasized above, loop corrections to the jet function have to be calculated
in the light-cone gauge. In this gauge the contributions of the diagrams $\{4
j\}$, where the gluon is radiated by the heavy quark,  vanishes. The reason is
that the heavy quark can radiate longitudinally polarized collinear gluons
only; the contribution of these gluons is excluded by the condition $A\cdot
n_+=0$. Diagrams $\{11\}$, $\{22\}$, $\{55\}$ do not contribute also because
they do not contain the hard-collinear scale. Nonzero contributions are given
by the following diagrams:
\begin{align}
J_{\{15\}}&=\left( C_{F}-\frac{C_{A}}{2}\right) \,\delta(v-u)\left( -\ld
+(3-2\,\ln u)\ls\right.\notag\\& \qquad \left.+3\,\ln u-{\ln}^{2}{u}-8+\frac{{%
\pi}^{2}}{6}\right),\\
J_{\{66\}}&=C_{A}\,\delta(v-u)\left( 2\ld-\left(\frac{11}{3}-4\,\ln
u\right)\ls-\frac{11\,\ln u}{3}\right.\notag\\& \qquad \left.
+2\,{\ln}^{2}{u}+\frac{67}{9}-%
\frac{{\pi}^{2}}{3}\right)+ T_F\,n_f \delta(v-u)
\left(\frac{4}{3}\,\ls+\frac{4}{3}\,\ln u-\frac{20}{9}
\right),\\
J_{\{16\}}&=\frac{C_{A}}{2}\,\delta(v-u)\left( -3\ld +\left(3-6\,\ln
u\right)\ls\right.\notag\\& \qquad \left.
+3\,\ln u-3\,{\ln}^{2}{u}-6+\frac{{\pi}^{2}}{2}%
\right)\,,\\
J_{\{33\}}&=2C_{F}~\bar{v}\,\left( \ls+\ln(v\,\bar
{v})+1\right)\,,\\
J_{\{23\}}&=2~\left( C_{F}-\frac{C_{A}}{2}\right) \,\left( \theta (v-\bar{u}%
)~\left( \,\frac{\,\bar{v}~\left( v-\bar{u}\right) }{v\,u}\left( \ls
+\ln (\bar{v})\right) \,\right. \right.  \nonumber\\
& \left. +\,\frac{\left( v-\bar{u}\right) }{\bar{u}}\ln v-\frac{\,\left( v-%
\bar{u}\right) ^{2}}{v\,u\,\bar{u}}\ \ln \left( 1-\frac{\bar{v}}{u}\right)
\,+\frac{\bar{v}}{v\,u}\right)  \nonumber\\
& \left. +\,\theta (\bar{u}-v)~\frac{\bar{u}-v}{\bar{u}}\,\left( \frac{1}{%
\epsilon }-\ln v\bar{v}\right) \,\right)\,.
\end{align}
\begin{align}
& J_{\{25\}}+J_{\{26\}}+J_{\{12\}}=\left(C_{F}-\frac{C_{A}}{2}\right) \nonumber \\
& \times \,2\,\left[ \frac{\theta (v-u)}{v-u}%
\,\left(\ls+\ln (v-u)\right) +\frac{\theta (u-v)}{u-v}%
\,\left( \ls+\ln (u-v)\right) \,\right] _{+} \nonumber \\
& +\left( C_{F}-\frac{C_{A}}{2}\right) \,\delta (v-u)\,2\,\left( \ld
+\ln(u\bar{u})\,\ls+\frac{{\ln }^{2}{\bar{u}}}{2%
}+\frac{{\ln }^{2}{u}}{2}+\frac{{\pi }^{2}}{6}\right)  \nonumber \\
& +C_{F}\,\theta (u-v)\,\frac{2\bar{v}}{\bar{u}\,u}\,\left( \ls
+\ln \left( \frac{v\,\left( u-v\right) }{u}\right) -\frac{\bar{u}%
}{\bar{v}}\right)  \nonumber \\
& -\left( C_{F}-\frac{C_{A}}{2}\right)\,\frac{2\,\theta (u-v)}{u}\left(
\frac{\bar{u}-v}{\bar{u}}\ls-\frac{u}{u-v}\ln \frac{v}{u}+\frac{%
\left( \bar{u}-v\right) }{\bar{u}}\,\ln \frac{v\,\left( u-v\right) }{u}%
-1\right)  \nonumber \\
& +\left( {C_{F}-}\frac{{C_{A}}}{2}\right)\,\frac{2\,\theta (v-u)}{v}\left(
-\ls+\frac{u}{v-u}\ln \frac{\bar{v}}{\bar{u}}-\ln (v-u)+%
\frac{\bar{v}}{\bar{u}}\right) \, \nonumber \\
& +\left( {C_{F}-}\frac{C_{A}}{2}\right) \,\frac{2\,\left( \bar{u}-v\right) }{%
\bar{u}}\,\,\left( \ls +\ln v\bar{v}\right) \, \nonumber \\
& -{C_{F}\,}\frac{2\,\bar{v}\,}{\bar{u}}\left( \ls+\ln v\bar{%
v}-\frac{\bar{u}}{\bar{v}}\right) -2\,{C_{F}}\,\frac{\,\theta (v-u)}{v} \,.
\end{align}
As in the case of the hard coefficient function, the corrections to the jet
function are finite in the $u\to 0$ limit. It in particular means that the
corresponding convolution integral with pion light-cone distribution amplitude
is convergent. The final result for the jet functions has the form:
\begin{multline}
J\left(u, v, \ln\frac{2E\omega}{\mu^2}, \alpha_s(\mu)\right)= \frac{\alpha_s}{u\, \omega}\left( \delta(u-v)\right.\\
\left.+\frac{\alpha_{s}}{4\pi}~\left( C_{F}~\ln^2\frac{2E\omega}{\mu^2}~\delta
(u-v)+j_{1}(u,v)~\ln\frac{2E\omega}{\mu^2}+j_{2}(u,v)\right)\right)\,,
\end{multline}
where functions $j_1(u,v)$ и $j_2(u,v)$ are defined by
\begin{align}
j_{1}(u,v) & =\delta(v-u)~\left( -\frac{C_{A}}{2}~\left( \frac{13}{3}-2\ln
u\right) +\left( C_{F}-\frac{C_{A}}{2}\right) ~\left( 3+2\ln\bar {u}\right)
+\frac{4}{3}\,T_F\,n_f\right) \nonumber \\
& -\left( \frac{{C_{A}}}{2}-{C_{F}}\right) ~2\,\left[ \frac{\theta (v-u)}{v-u%
}\,+\frac{\theta(u-v)}{u-v}\right] _{+} \nonumber \\
& +\left( C_{F}-\frac{C_{A}}{2}\right) \,\frac{2~\left( v-\bar{u}\right) }{%
\bar{u}}\left( \theta(v-\bar{u})~\frac{\bar{v}~\bar{u}}{v\,u}+\,\theta (\bar{%
u}-v)~\,\right) \nonumber \\
& -\left( C_{F}-\frac{C_{A}}{2}\right) \,\,2\left( ~\theta(u-v)~\frac {\bar{u%
}-v}{u\bar{u}}+\theta(v-u)~\frac{1\,}{v}-\frac{\bar{u}-v}{\bar{u}}\,\right)
\nonumber \\
& +{C_{F}~2}\left( -\frac{u\,\bar{v}\,}{\bar{u}}+~\theta(u-v)\,\frac{\bar{v}%
}{\bar{u}\,u}\,\,\right)\,,
\end{align}
\begin{align}
j_{2}(u,v)& =\delta (v-u)\left( \frac{C_{A}}{2}\left( \frac{80}{9}-\frac{\pi ^{2}%
}{6}-\frac{13}{3}\ln u+\ln ^{2}u\right) - T_F\,n_f
\left(\frac{20}{9}-\frac{4}{3}\,\ln u
\right)\right.  \nonumber \\
& \left. +\left( C_{F}-\frac{C_{A}}{2}\right) \left( -8+\frac{\pi ^{2}}{2}%
+\ln ^{2}\bar{u}+3\ln u\right) \right)  \nonumber \\
& +\left({C_{F}}- \frac{{C_{A}}}{2}\right) ~2\,\left[ \frac{\theta (v-u)}{v-u%
}\,\ln (v-u)+\frac{\theta (u-v)}{u-v}\,\ln (u-v)\,\right] _{+} \nonumber \\
& -2\,{C_{F}}~\left( \frac{\,\theta (v-u)}{v}+\frac{\theta (u-v)}{u}\right)
\nonumber \\
& +\left( C_{F}-\frac{C_{A}}{2}\right) \,\,\theta (u-v)~\frac{2}{u}\left(
\frac{v}{u-v}\ln \frac{v}{u}-\ln \left( u-v\right) +1\right)  \nonumber \\
& +\left( {C_{F}-}\frac{{C_{A}}}{2}\right) ~\theta (v-u)~\frac{2\,}{v}\left(
\frac{u}{v-u}\ln \frac{\bar{v}}{\bar{u}}-\ln (v-u)+\frac{\bar{v}}{\bar{u}}%
\right) \, \nonumber \\
& +\left( C_{F}-\frac{C_{A}}{2}\right) \,\frac{2}{vu}~\theta (v-\bar{u}%
)~\left( \bar{v}-\frac{\,\left( v-\bar{u}\right) ^{2}}{\bar{u}}\ \ln \frac{%
\bar{v}~\left( u-\bar{v}\right) }{u}\right) \,\, \nonumber \\
& +{C_{F}~2}\left( 1+{\bar{v}}-\frac{{\bar{v}}u\,}{\bar{u}}\ln v\bar{v}%
\right) +2~\frac{\theta (u-v)}{u\bar{u}}\ln \frac{v\,\left( u-v\right) }{u}%
\left( C_{F}~-\frac{C_{A}}{2}\,\,v\right)\,.
\end{align}
The result obtained is in complete agreement with the result calculated in
\cite{Neubert1}\footnote{According to a private communication from M. Beneke,
his result for the loop corrections to the jet function, which has been
obtained in cooperation with D. S. Yang, coincides with the result calculated
in \cite{Neubert1}.} (In our notations the corresponding correction is
$-j_\|(\bar{u},\bar{v})$, where $j_\|$ is the function derived in
\cite{Neubert1})


\subsection{Collinear region}

\begin{figure}
[hptb]
\begin{center}
\
\includegraphics[width=.15\textwidth]{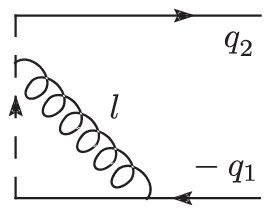}%
\hspace{1cm}
\includegraphics[width=.15\textwidth]{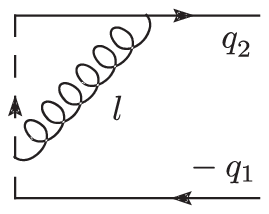}%
\hspace{1cm}
\includegraphics[width=.15\textwidth]{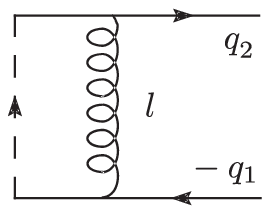}%
\\
\parbox[t]{.15\textwidth}{\caption{}\label{f_LA1}}
\hspace{1cm}
\parbox[t]{.15\textwidth}{\caption{}\label{f_LA2}}
\hspace{1cm}
\parbox[t]{.15\textwidth}{\caption{}\label{f_LA3}}
\end{center}
\end{figure}
In this Section we consider factorization of the collinear singularities, i.e.,
the singularities associated with the radiation of a collinear gluon (momentum
scaling is $(\lambda^4,\lambda^2,1))$. Such singularities have an origin in the
singularity of a collinear quark propagator when a gluon is radiated in the
direction of the quark. These singularities can be regularized by introduction
of the soft transverse momentum (see (\ref{qperp1}), (\ref{qperp2})). The
logarithms that result are $\ln \mu^2/\mathbf{q}^2_\perp$, where $\mu$ is a
factorization scale. As we shall see later, all collinear divergencies
correspond to the renormalization of the pion LCDA. Thus, these singularities
are cancelled by the substraction of the loop corrections corresponding to
renormalization of the LCDA from the total loop corrections:
\begin{align}
\Delta f&=\Phi^+(\omega)\otimes T_0(\omega,u')\otimes \delta \phi_{\pi}(u')&\\
\delta \phi_\pi(u',\mu)&=\frac{g^2}{(4\pi)^2}\,C_F\,
\ln\frac{\mu^2}{\mathbf{q}_\perp^2}\:V^{(1)}_{\pi}(u',u)\otimes\phi_\pi(u,\mu)\,,&
\end{align}
where $V^{(1)}_{\pi}(u',u)$ is ERBL kernel \cite{ER}, \cite{BL},
\begin{align}
\left\langle \pi(q)\left\vert\,\bar{q}_{\alpha}(sn_{+})W\left(s n_{+},0\right)
q_{\beta}(0)\right\vert 0\right\rangle =\frac{if_{\pi}}{4}\left(  n_{+}\cdot
q\right)  ~\left[  \frac{\not\!n}{2}\gamma_{5}\right]  _{\beta\alpha}\int
_{0}^{1}du\,e^{i u s \left(  n_{+}\cdot q\right)  }\phi_{\pi}(u)\,.
\label{piLCDA}
\end{align}
The contribution of each diagram for both the collinear and soft regions is
denoted by
\begin{align}
\Delta f =-K_{\pi}\,K_{B}\,C_{F}\,N_{c}\,(D-2)\,\frac{g^{2}}{E^{2}}\int_0^1
\frac{du}{u}\int_0^\infty \frac{d\omega}{\omega}\,\Phi_+(\omega)
A_{\{ij\}}(\omega,u) \phi_\pi(u) \,,
\end{align}
where $i$ and $j$ are the numbers of positions in the Figure \ref{jnum} for the
collinear gluon emission and absorbtion (see also definitions for the
tree-level amplitude (\ref{sp2})).

First of all, we consider the contributions of the diagrams where the light
antiquark radiates the collinear gluon, i.e., the $\{1i\}$ diagrams. Using the
following notation for the amplitudes~$A_{\{1i\}}$
\begin{equation}
A_{\{1i\}}=2i\,g^2\int [dl]\,\frac{\tilde{A}_{\{1i\}}}{\left( l^{2}+i0\right)
\left( (q_{1}-l)^{2}+i0\right)}\,,
\end{equation}
we find the following integrands
\begin{align}
\tilde{A}_{\{15\}}&=\left( C_{F}-\frac{C_{A}}{2}\right) \left(
-1+\frac{u}{\beta}\right), & \tilde{A}_{\{16\}}&=-\frac{C_{A}}{2}\,,
\\
\tilde{A}_{\{14\}}&=\left(C_{F}-\frac{C_{A}}{2}\right) \left(
-\frac{u}{\beta}+\frac{u}{\beta -1}\right), & \tilde{A}_{\{13\}}&=\left(
C_{F}-\frac{C_{A}}{2}\right) \left(-\frac{u}{\beta -1}\right)\,,
\end{align}
where $\beta=(l\cdot n_+)/2E$ is a dimensionless Sudakov parameter. The
corresponding corrections to the pion LCDA is depicted in Figure \ref{f_LA1}.
The dash line denotes the Wilson line $W\left(s n_{+},0\right)$ which is
connecting the quark fields in the LCDA. The sum of the contributions of the
\{1i\} diagrams is cancelled by the correction convoluted with the tree-level
coefficient function $1/u$:
\begin{align}
& \int_0^1 \frac{du}{u}\,\phi_\pi(u) \sum_i A_{\{1i\}}(u) \rightarrow
\int_{0}^{1}\frac{du^{\prime}}{u^{\prime}}\int_{0}^{1}du~V_{\ref{f_LA1}}(u^{\prime
},u)~\phi_{\pi}(u)\\
& =-2ig^{2}C_{F}\int_{0}^{1}\frac{du^{\prime}}{u^{\prime}}\int_{0}^{1}%
\phi_{\pi}(u)~du\int[dl]~\frac{~\beta-u}{\left(  (l-q_{1})^{2}+i0\right) \left(
l^{2}+i0\right)  }~\frac{\delta(u^{\prime}-u)-\delta\left(  u^{\prime
}-(u-\beta)\right)  }{\beta+i0}\nonumber\\
& =-2ig^{2}\int_{0}^{1}\frac{\phi_{\pi}(u)}{u}~du\int[dl]~\frac{C_{F}}{\left(
(l-q_{1})^{2}+i0\right)  \left(  l^{2}+i0\right)  }\,.%
\end{align}
The collinear singularities corresponding to the radiation of the collinear
gluon by the light quark (the \{2i\} diagrams) can be considered in a similar
manner:
\begin{align}
A_{\{2i\}}=2i\,g^2\int [dl]\,\frac{\tilde{A}_{\{2i\}}}{\left( l^{2}+i0\right)
\left( (q_{2}-l)^{2}+i0\right)}\,,
\end{align}
\begin{align}
\tilde{A}_{\{25\}} &=\left( C_{F}-\frac{C_{A}}{2}\right) \left(
\frac{\beta -\bar{u}}{\beta }-\frac{\beta -\bar{u}}{\beta +u}%
\right),  & \tilde{A}_{(26)}  &=-\frac{C_{A}}{2}\left( \frac{\beta -\bar{u}%
}{\beta +u}\right),\\
\tilde{A}_{\{24\}} &=\left( C_{F}-\frac{C_{A}}{2}\right) \left( -%
\frac{\beta -\bar{u}}{\beta }+\frac{\beta -\bar{u}}{\beta -1}\right),
& \tilde{A}_{\{23\}}  &=\left( C_{F}-\frac{C_{A}}{2}\right) \left( -%
\frac{\beta -\bar{u}}{\beta -1}\right).
\end{align}
The sum of the contributions of these diagrams is cancelled by the correction
depicted in Figure~\ref{f_LA2}.
\begin{align}
& \int_0^1 \frac{du}{u}\,\phi_\pi(u)\sum_i A_{\{2i\}}(u)\rightarrow
\int_{0}^{1}\frac{du^{\prime}}{u^{\prime}}\int_{0}^{1}du~V_{\ref{f_LA2}}(u^{\prime
},u)~\phi_{\pi}(u)\\
& =  2ig^{2}C_{F}\int_{0}^{1}\frac{du^{\prime}}{u^{\prime}}\int_{0}^{1}%
\phi(u)~du\int[dl]~\frac{~(\bar{u}-\beta)}{\left(  (l-q_{2})^{2}+i0\right)
\left(  l^{2}+i0\right)  }~\frac{\left(  \delta(u^{\prime}-u)-\delta\left(
u^{\prime}-(u+\beta)\right)  \right)  }{\beta+i0}\nonumber\\
& = 2ig^{2}\int_{0}^{1}\frac{\phi(u)}{u}~du\int[dl]~\frac{C_{F}}{\left(
(l-q_{2})^{2}+i0\right)  \left(  l^{2}+i0\right)  }~\frac{\bar{u}-\beta
}{u+\beta}\,.%
\end{align}
The diagram \{12\} with the collinear gluon exchange between the quark and
antiquark
\begin{align}
A_{\{12\}}= -2ig^2\,C_{F}\,\int [dl] \frac{\left( \mathbf{l}_{\perp
}-\mathbf{q}_{\perp }\right) ^{2}} {\left( l^{2}+i0\right)
\left((q_{1}-l)^{2}+i0\right) \left( (q_{2}+l)^{2}+i0\right) }\frac{u}{\beta
-u+i0}
\end{align}
is cancelled by the corresponding diagram Fig.~\ref{f_LA3}. The self-energy
diagrams $\{11\}$ and $\{22\}$ are trivially cancelled by the renormalization
of the quark fields in the pion light-cone distribution amplitude.


\subsection{Soft region}

\begin{figure}
[hptb]
\begin{center}
\includegraphics[width=.15\textwidth]{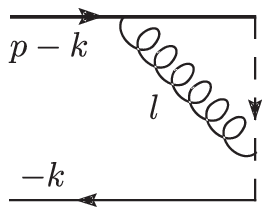}%
\hspace{1.5cm}
\includegraphics[width=.15\textwidth]{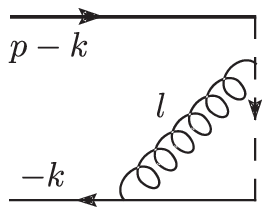}%
\hspace{1.5cm}
\includegraphics[width=.15\textwidth]{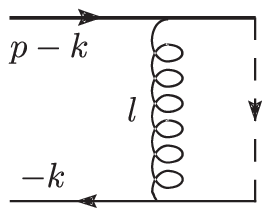}%
\\
\parbox[t]{.15\textwidth}{\caption{}\label{f_LB1}}
\hspace{1.5cm}
\parbox[t]{.15\textwidth}{\caption{}\label{f_LB2}}
\hspace{1.5cm}
\parbox[t]{.15\textwidth}{\caption{}\label{f_LB3}}
\end{center}
\end{figure}

Below is a consideration of the singularities associated with the soft region
$l\sim (\lambda^2,\lambda^2,\lambda^2)$. This region occurs in the diagrams
with a gluon radiated  by the slow-moving heavy quark or by the soft spectator.
Assuming $p^2=m_b^2$, the singularities arise as logarithms $\ln \mu^2/\absk$.
These "big"\ logarithms are absorbed by the renormalization of the $B$-meson
light-cone distribution amplitude:
\begin{align}
\Delta f&=\delta \Phi^+(\omega')\otimes T_0(\omega',u)\otimes \phi_{\pi}(u)&\\
\delta \Phi^+(\omega',\mu)&=\frac{g^2}{(4\pi)^2}\,C_F\,\left(
\Gamma^{(1)}\,\ln^2\frac{\mu^2}{\absk}\,\Phi^+(\omega',\mu)+
\ln\frac{\mu^2}{\absk}\,U^{(1)}(\omega',\omega)\otimes\Phi^+(\omega,\mu)\right)\,,&
\end{align}
where $\Gamma^{(1)}$ and $U^{(1)}(\omega',\omega)$ are the functions contained
in the kernel of the evolution equation \cite{Bosh}, \cite{NL} of the
corresponding B-meson light-cone distribution amplitude $\Phi^+(\omega)$
\cite{GN}, \cite{Beneke0}
\begin{multline}
\left\langle 0\left\vert \bar{q}_{\alpha}(sn)~W(sn,0)~h_{v\beta
}(0)\right\vert \bar{B}(p)\right\rangle \\
=-\frac{i f_{B}M_{B}}{4}\int_{0}^{\infty}d\omega\,e^{-i\omega s}\left(
\frac{1+\not\!v}{2}\left[  \left(  n\cdot v\right) \not\!n_{+}\,\Phi_{+}%
(\omega)+\left(  n_{+}\cdot v\right)
\not\!n\,\Phi_{-}(\omega)\right]\gamma_5\right)
_{\beta\alpha}\,.\label{BLCDA}%
\end{multline}

First of all we consider the diagrams with a gluon radiated by the heavy quark.
As in the case of the collinear region, we introduce the following notation:
\begin{equation}
A_{\{4i\}}=2i\,g^2\int [dl] \frac{m_b~\tilde{A}_{\{4i\}}}{\left(
l^{2}+i0\right) \left( (p-k-l)^{2}-m_b^{2}\right) }\,.
\end{equation}
The integrands of the \{4i\} diagrams are as follows:
\begin{equation}
\tilde{A}_{\{42\}}=\left( C_{F}-\frac{C_{A}}{2}\right) \frac{1}{\alpha
}\,,\quad \tilde{A}_{\{41\}}=\left(
C_{F}-\frac{C_{A}}{2}\right)\left(\frac{1}{\alpha +\omega}-\frac{1}{\alpha
}\right)\,,\quad \tilde{A}_{\{46\}}=\frac{C_{A}}{2}\frac{1}{\alpha+\omega }\,,
\end{equation}
where $\alpha=(l\cdot n)$. Being convoluted with the tree-level coefficient
function~$1/\omega$\ , the contribution to the LCDA's renormalization
(Fig.\ref{f_LB1}) with gluon radiation by the heavy quark cancels the sum of
the contributions of the \{4i\} diagrams
\begin{align}
& \int_0^\infty\frac{d\omega}{\omega}\,\Phi_+(\omega) \sum_i A_{\{4i\}}(\omega)
\rightarrow
\int_{0}^{\infty}\frac{d\omega^{\prime}}{\omega^{\prime}}\int_{0}^{\infty
}d\omega\,V_{\ref{f_LB1}}(\omega^{\prime},\omega)\,\Phi_{+}(\omega)\\
& =-ig^{2}C_{F}\int_{0}^{\infty}\frac{d\omega^{\prime}}{\omega^{\prime}}%
\int_{0}^{\infty}d\omega\int[dl]\,\frac{1}{\left(  (l+k)\cdot v-i0\right)
\left( l^{2}+i0\right)  }\,\frac{\delta(\omega^{\prime}-\omega)-\delta
(\omega^{\prime}-\omega-\alpha)}{\alpha-i0}\,\Phi_{+}(\omega)\nonumber\\
& =-ig^{2}\int_{0}^{\infty}\frac{d\omega}{\omega}\,\Phi_{+}(\omega
)\int[dl]\,\frac{C_{F}}{\left(  (l+k)\cdot v-i0\right)  \left(  l^{2}%
+i0\right)}\,\frac{1}{\omega+\alpha}\,.%
\end{align}
The contributions of the diagrams with the gluon radiation by the soft
spectator can be considered in a similar way
\begin{equation}
A_{\{5i\}}= -2i\,g^2\int [dl]\frac{\tilde{A}_{\{5i\}}}{%
\left( (l-k)^{2}+i0\right) \left( l^{2}+i0\right) }\,,
\end{equation}
\begin{equation}
\tilde{A}_{\{51\}}=\left( C_{F}-\frac{C_{A}}{2}\right) \frac{\alpha -\omega
}{\alpha }\,,\quad \tilde{A}_{\{52\}}=\left( C_{F}-\frac{C_{A}}{2}\right)
\frac{\omega }{\alpha }\,,\quad \tilde{A}_{\{56\}}=\frac{C_{A}}{2}\,.
\end{equation}
The sum of the contributions is cancelled by the contribution to the LCDA's
renormalization (Fig. \ref{f_LB2}) with radiation of a gluon by the soft
spectator
\begin{align}
& \int_0^\infty\frac{d\omega}{\omega}\,\Phi_+(\omega) \sum_i A_{\{5i\}}(\omega)
\rightarrow
\int_{0}^{\infty}\frac{d\omega^{\prime}}{\omega^{\prime}}\int_{0}^{\infty
}d\omega~V_{\ref{f_LB2}}(\omega^{\prime},\omega)~\Phi_{+}(\omega)\\
& =-2ig^{2}C_{F}\int_{0}^{\infty}\frac{d\omega^{\prime}}{\omega^{\prime}}%
\int_{0}^{\infty}d\omega\int[dl]\,\frac{(\alpha-\omega)}{\left(
(l-k)^{2}+i0\right)  \left(  l^{2}+i0\right)  }\,\frac{\delta(\omega^{\prime
}-\omega)-\delta(\omega^{\prime}-(\omega-\alpha))}{\alpha-i0}\,\Phi_{+}%
(\omega)\nonumber\\
& =-2ig^{2}\int_{0}^{\infty}\frac{d\omega}{\omega}\,\Phi_{+}(\omega
)\int[dl]\,\frac{C_{F}}{\left(  (l-k)^{2}+i0\right)  \left(  l^{2}+i0\right)
}\,.
\end{align}
The situation with the diagrams $\{44\}$ and $\{55\}$ is similar to the one in
the collinear region,  these diagrams correspond to the renormalization of the
quark field in the light-cone distribution amplitude of the $B$-meson. The soft
region of the diagram $\{45\}$ is power suppressed; it exactly corresponds to
the absence of the contribution to the evolution kernel from the diagram
Fig.~\ref{f_LB3}.


\subsection{Soft-collinear}

In this Section we consider the contribution of the soft-collinear region,
i.e., the region where the momentum of a gluon has its' components
$\displaystyle l=\left((l\cdot n), l_\perp, (l\cdot n_+)\right)$ that are of
order $(\lambda^4, \lambda^3, \lambda^2)$.

According to the idea of the method of expanding by regions, for the difference
of the form factors to be independent on the intermediate factorization scale,
it is necessary to sum the contributions of \textit{all} the regions, the
soft-collinear region included. However, the contribution of this region does
not correspond to the summation of the leading logarithms for any of the
objects in the expression (\ref{factor}), therefore, it has to be cancelled in
the sum of the diagrams by itself. This cancellation corresponds to the idea
that, in the framework of the effective theory, after integration out of the
hard and hard-collinear modes the soft and collinear degrees of freedom should
be decoupled from one another. At this specified condition, a matrix element,
containing the soft and collinear modes, can be factorized into matrix elements
of "soft"\ and of "collinear"\ operators separately. This is of critical
importance for a derivation of a factorization formula of type~(\ref{factor}).
We show the cancellation of soft-collinear region contributions for pion final
state, though it is liable to brake down in general case \cite{messenger}.

\begin{figure}
[hptb]
\begin{center}
\includegraphics[width=.6\textwidth]{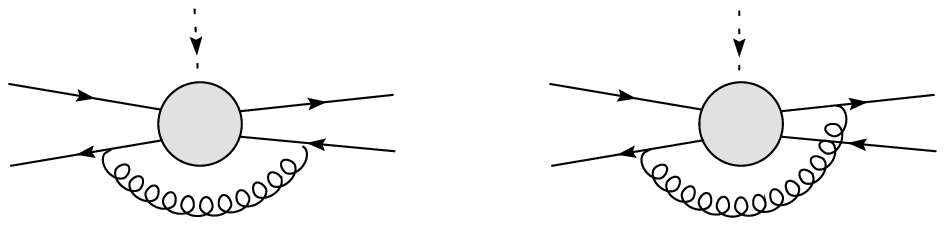}%
\\
\parbox[t]{.15\textwidth}{\caption{}\label{f_SC1}}
\end{center}
\end{figure}

The cancellation of the contribution of the soft-collinear region occurs in
line with the idea of \textit{color transparency} \cite{Bj}, which is
underlying the factorization approach to exclusive processes: the final state
hadron produced by hard scattering is a color singlet state. Thus, there is a
rather weak (dipole) interaction with gluons, the wavelength of which is larger
than the size of the hadron. Therefore, it is hoped that there is a
cancellation of the soft logarithmic singularities of the radiative corrections
between graphs in which the soft gluon couples to a parton in the initial state
and different constituents of the final state (see Fig.~\ref{f_SC1}). Whereas
the remaining infrared singularities factorize into the LCDAs of each hadron
separately.

To calculate loop corrections it is convenient to introduce the followng
dimensionless variables:
\[
(l\cdot n_+) \rightarrow\frac{\absk}{\omega}~\beta,~(l\cdot n) \rightarrow
\frac{\mathbf{q}_{\perp}^{2}}{2E}~\alpha,~\mathbf{l}_{\perp}^{2}%
\rightarrow\frac{\absk\mathbf{q}_{\perp}^{2}}{2E\omega}%
~\mathbf{l}_{\perp}^{2}.%
\]
The contribution of the diagram \{25\} with these variables takes the form:
\begin{align}
\{25\}  =i\,g^2 \left(  C_{F}-\frac{C_{A}}{2}\right) \left(  \frac{\mu^{2}2E\omega}{\absk%
\mathbf{q}_{\perp}^{2}}\right)  ^{\epsilon} \int[dl]~\frac{2\bar{u}}{\left(
l^{2}+i0\right)  (\beta-1+i0)(\alpha \bar{u}-1+i0)}.
\end{align}
After the additional substitution $\alpha\to\alpha/\bar{u}, l_\perp\to
l_\perp/\bar{u}^{1/2}$ we find
\begin{align}
\{25\}  & =i\,g^2 \left(  C_{F}-\frac{C_{A}}{2}\right) \left(  \frac{\mu^{2}2E\omega}{\absk%
\mathbf{q}_{\perp}^{2}}\right)  ^{\epsilon}  \int[dl]~\frac{2\bar{u}^\epsilon%
}{\left(  l^{2}+i0\right)  (\beta-1+i0)(\alpha-1+i0)}\,. \label{sc_25}%
\end{align}
The final result of the integral is the expression (\ref{sc}). The similar
expression for the diagram \{15\} is
\begin{align}
\{15\}  =-i\,g^2 \left(  C_{F}-\frac{C_{A}}{2}\right) \left(  \frac{\mu^{2}2E\omega}{\absk%
\mathbf{q}_{\perp}^{2}}\right)  ^{\epsilon} \int[dl]~\frac{2u^\epsilon}{\left(
l^{2}+i0\right)  (\beta-1+i0)(\alpha - 1+i0)}\,.\label{sc_15}
\end{align}
Contrary to our expectations, the expressions (\ref{sc_25}) and (\ref{sc_15})
are not only of opposite sign, but they also differ by the factors $u^\epsilon$
and $\bar{u}^\epsilon$. It immediately follows that in the case of the partons
with $u$ and $\bar{u}$ fractions of the longitudinal momentum there is a
cancellation of double Sudakov logarithms only, but not of single ones.
However, the pion LCDA is symmetric with a substitution $u\leftrightarrow
\bar{u}$ while the sum of the expressions (\ref{sc_25}) and (\ref{sc_15}) is
antisymmetric. Therefore, being averaged over the pion light-cone distribution
amplitude, the contribution of the soft-collinear region vanishes.

The diagrams \{14\} and \{24\} can be considered in a similar manner:

\begin{align}
\{14\}  & =- i\,g^2 \left(
C_{F}-\frac{C_{A}}{2}\right) \left(  \frac{\mu^{2}2E\omega}{\absk%
\mathbf{q}_{\perp}^{2}}\right)
^{\epsilon}\int[dl]~\frac{2u~\absk}{l^{2}(1+\alpha u)\left(
\absk\bar{\beta}-\omega^{2}\right)  }\label{sc_14}\,,\\
\{24\}  & =i\,g^2 \left( C_{F}-\frac{C_{A}}{2}\right) \left(  \frac{\mu^{2}2E\omega}{\absk%
\mathbf{q}_{\perp}^{2}}\right)  ^{\epsilon}
\int[dl]~\frac{2\bar{u}~\absk}{l^{2}(1+\alpha\bar{u})\left(
\absk \bar{\beta}-\omega^{2}\right)  }\,.\label{sc_24}%
\end{align}

Even before the integration one can see that the sum of the expressions
(\ref{sc_14}), (\ref{sc_24}) is antisymmetric with respect to the substitution
$u\leftrightarrow \bar{u}$. Hence, it vanishes in the convolution with the pion
light cone-distribution amplitude.


\section{Conclusion}
The semileptonic decay $B \to \pi l \nu$ has been discussed in this article. It
is known that at a large energy of recoil to the lepton pair there is the
relation of the form factors that parametrize an amplitude of this decay. This
relation can be violated by the hard and semihard radiative corrections. In
this paper we have studied the factorizable contribution of these corrections.
In particular, loop corrections to the hard Wilson coefficient and to the jet
function have been calculated. In contrast to the soft-collinear effective
theory, to calculate loop corrections we have used the method of expanding by
regions. The corrections obtained are in complete agreement with previous
results of Beneke, Kiyo and Yang \cite{Beneke1} and results of Hill, Becher,
Lee and Neubert \cite{Neubert1}. The factorization of the soft and collinear
singularities has been shown at one-loop level.  We have demonstrated the
cancellation of the contribution of the soft-collinear modes by taking into
account the symmetry of the pion light-cone distribution amplitude; this fact
is of critical importance to the factorization approach to this decay.

\vspace{.3cm}

\noindent \textit{Note added.} After I completed this paper, the results of M.
Beneke and D. Yang mentioned above have appeared in preprint \cite{Beneke3}.

\vspace{.5cm}

\noindent I would like to thank N. Kivel for his helpful comments and
discussions and M. Beneke for pointing out the dependence of the one-loop jet
function on the choice of evanescent operators. The investigation was supported
by the Russian Foundation for Basic Research through Grant No. 05-02-16627-a
and by Institut $\mathrm{f\ddot{u}r}$ Theoretische Physik-II
Ruhr-$\mathrm{Universit\ddot{a}t}$ through\\ Graduiertenkolleg "Physik der
Elementarteilchen an Beschleunigern und im Universum".
\bibliographystyle{amsplain}

\end{document}